\magnification\magstephalf
\overfullrule 0pt

\font\rfont=cmr9 at 9 true pt
\def\ref#1{$^{\hbox{\rfont {[#1]}}}$}

  %%Fonts

%%%family 8%%%
\font\tenbfit=cmbxti10
\font\sevenbfit=cmbxti10 at 7pt
\font\fivebfit=cmbxti10 at 5pt
\newfam\bfitfam 
\textfont\bfitfam=\tenbfit  \scriptfont\bfitfam=\sevenbfit
\scriptscriptfont\bfitfam=\fivebfit

%%%family 9%%%
\font\tenbit=cmmib10
\newfam\bitfam
\textfont\bitfam=\tenbit%

%%%family 10%%%
\font\tenmbf=cmbx10
\font\sevenmbf=cmbx7
\font\fivembf=cmbx5
\newfam\mbffam
\textfont\mbffam=\tenmbf \scriptfont\mbffam=\sevenmbf
\scriptscriptfont\mbffam=\fivembf

%%%family 11%%%
\font\tenbsy=cmbsy10
\newfam\bsyfam 
\textfont\bsyfam=\tenbsy%
\font\fourteenbfit=cmbxti12 scaled\magstep1

\font\fourteenbf=cmbx12 scaled\magstep1

\font\tenbfit=cmbxti10
\font\sevenbfit=cmbxti10 at 7pt
\font\fivebfit=cmbxti10 at 5pt
\newfam\bfitfam 
\textfont\bfitfam=\tenbfit  \scriptfont\bfitfam=\sevenbfit
\scriptscriptfont\bfitfam=\fivebfit

\font\eightit=cmti8

  %%Greek
   
\def\e{\epsilon}

\def\pmb#1{\setbox0=\hbox{#1}% \kern-.025em\copy0\kern-\wd0
 \kern.05em\copy0\kern-\wd0 \kern-.025em\raise.0433em\box0 }

\def\slash{/\kern-.5em}

  %%Fractions
\def \half {{\textstyle {1 \over 2}}}

 %

 %%FORMATTING

\def\boxit#1{\vbox{\hrule\hbox{\vrule\kern1pt\vbox
{\kern1pt#1\kern1pt}\kern1pt\vrule}\hrule}}

\def\h{\hfill\break}
\parskip=6pt
\parindent=0pt
\hsize=17truecm\hoffset=-5truemm
\vsize=24truecm
\def\footnoterule{\kern-3pt
\hrule width 17truecm \kern 2.6pt}

  %%REFERENCES
%     \defref\label{text}
% generates a number, assigns it to \label, generates an entry.
% To list the refs,  \listrefs
% (Extracted and adapted from harvmac.tex by P Ginsparg)

\catcode`\@=11 % This allows us to modify PLAIN macros.

\def\nolabels{\def\wrlabeL##1{}\def\eqlabeL##1{}\def\reflabeL##1{}}
\def\writelabels{\def\wrlabeL##1{\leavevmode\vadjust{\rlap{\smash%
{\line{{\escapechar=` \hfill\rlap{\sevenrm\hskip.03in\string##1}}}}}}}%
\def\eqlabeL##1{{\escapechar-1\rlap{\sevenrm\hskip.05in\string##1}}}%
\def\reflabeL##1{\noexpand\llap{\noexpand\sevenrm\string\string\string##1}}}
\nolabels
\global\newcount\refno \global\refno=1
\newwrite\rfile
\def\defref{$^{{\hbox{\rfont [\the\refno]}}}$\nref}
\def\nref#1{\xdef#1{\the\refno}\writedef{#1\leftbracket#1}%
\ifnum\refno=1\immediate\openout\rfile=refs.tmp\fi
\global\advance\refno by1\chardef\wfile=\rfile\immediate
\write\rfile{\noexpand\item{#1\ }\reflabeL{#1\hskip.31in}\pctsign}\findarg}
%	horrible hack to sidestep tex \write limitation
\def\findarg#1#{\begingroup\obeylines\newlinechar=`\^^M\pass@rg}
{\obeylines\gdef\pass@rg#1{\writ@line\relax #1^^M\hbox{}^^M}%
\gdef\writ@line#1^^M{\expandafter\toks0\expandafter{\striprel@x #1}%
\edef\next{\the\toks0}\ifx\next\em@rk\let\next=\endgroup\else\ifx\next\empty%
\else\immediate\write\wfile{\the\toks0}\fi\let\next=\writ@line\fi\next\relax}}
\def\striprel@x#1{} \def\em@rk{\hbox{}} 
\def\lref{\begingroup\obeylines\lr@f}
\def\lr@f#1#2{\gdef#1{\defref#1{#2}}\endgroup\unskip}
\newwrite\lfile
{\escapechar-1\xdef\pctsign{\string\%}\xdef\leftbracket{\string\{}
\xdef\rightbracket{\string\}}}

\def\writestop{\def\writestoppt{\immediate\write\lfile{\string\p
ageno%
\the\pageno\string\startrefs\leftbracket\the\refno\rightbracket%
\string\def\string\secsym\leftbracket\secsym\rightbracket%
\string\secno\the\secno\string\meqno\the\meqno}\immediate\closeout\lfile}}
\def\writestoppt{}\def\writedef#1{}
\catcode`\@=12 % at signs are no longer letters

\rightline{DAMTP-1998-34}
\line{{\fourteenbf }\hfil M/C TH 98-9}

\bigskip\bigskip
\centerline{\fourteenbf SMALL {\fourteenbfit x}\ : TWO POMERONS!}
\vskip 10mm
\centerline{A Donnachie}
\vskip 3mm
\centerline{Department of Physics and Astronomy, University of Manchester,
Manchester M13 9PL}
\vskip 6mm
\centerline{P V Landshoff}
\vskip 3mm
\centerline{DAMTP, University of Cambridge, Cambridge CB3 9EW}
\vskip 10mm
\footnote{}{ad@a3.ph.man.ac.uk\ \ \ pvl@damtp.cam.ac.uk}
\midinsert\leftskip 8mm\rightskip 8mm
{\bf{Abstract}}

Regge theory provides a very simple and economical description of 
data for (i) the proton structure function with $x<0.07$ and all
available $Q^2$ values, (ii) the charm structure function, and (iii)
$\gamma\, p\to J/\psi\, p$.
The data are all in agreement with the assumption
that there is a second pomeron, with intercept about $1.4\,$.
They suggest also that the contribution from the soft pomeron is higher twist.\h
This means that there is  an urgent need to make perturbative evolution
compatible with Regge theory at small and not-so-small $x$, and to reassess
the magnitude of higher-twist contributions at quite small $x$.
\endinsert
\vskip 15mm

Regge theory is one of the great truths of particle physics\defref\regge{
P D B Collins, {\it Introduction to Regge Theory and High Energy Physics},
Cambridge University Press (1977)
}.
It is supposed to be applicable if $W^2$ is much greater than all
the other variables.
Thus, we expect it to be valid when $x$ is small enough,
even for large values of $Q^2$. Fits to large-$Q^2$ data have traditionally
concentrated on perturbative evolution; in this paper we report fits that
rather emphasise the Regge behaviour. Regge behaviour is not a substitute
for perturbative evolution, but a constraint on it, and an important task 
will be to understand better how the two coexist. Somewhat surprisingly,
we find that Regge behaviour is compatible with the data even for values
of $x$ up to $0.07$ or higher.

Regge theory involves pomerons.
Until HERA measurements of the proton structure function\defref\hera{ 
H1: C Adloff et al, Nuclear Physics B497 (1997)3\h
ZEUS: J Breitweg et al, Physics Letters B407 (1997) 432
} at  very small-$x$, and of $J/\psi$ photoproduction\defref\jpsi{
ZEUS: J Breitweg et al, Z Physik C75 (1997) 215 
}, 
it seemed consistent to suppose\defref\sigtot{
A Donnachie and P V Landshoff, Physics Letters B296 (1992) 227
}
that a single nonperturbative pomeron  with intercept close to 1.08
describes the whole of diffractive physics, including\defref\small{
A Donnachie and P V Landshoff, Z Physik C61 (1994) 139
}
the NMC and E665
structure function data at fairly small $x$. The more violent behaviour
observed at HERA calls rather for an intercept closer to 1.5 than 1.0,
and there has been much discussion of what is responsible for this.

One view\defref\capella{
A Capella, A Kaidalov, C Merino and J tran Thanh Van, Physics Letters B337
(1994) 358\h
M Bertini, M Giffon and E Predazzi, Physics Letters B349 (1995) 561
}
is that the pomeron intercept varies with $Q^2$, but we find this difficult to
believe. Regge theory relates\ref{\regge} the $W$-dependence 
of a process to the positions of singularities in
the complex angular momentum plane, and it does not naturally accommodate the
notion that these positions vary with $Q^2$.  The relative weights of the
contributions from these singularities can, however, vary with $Q^2$,
and so the effective power resulting from combining them can be
$Q^2$-dependent. 
We find less than persuasive the hypothesis\defref\gotsman{
E Gotsman, E M Levin and U Maor, Physical Review D49 (1994) 4321
}
that in soft processes there is a significant amount of shadowing, so that
the observed effective intercept of 1.08 actually is the result of the
pomeron having a significantly larger intercept, and that the rapid rises
observed at HERA are merely the result of the disappearance of the shadowing.
If there is a significant amount of shadowing, the apparent intercept should
be process-dependent, but all soft processes are found\defref\pdg{
Particle Data Group, Physical Review D54 (1996), page 191
}, to a remarkable accuracy,  to have the same value.

In this paper, therefore, we explore the notion that there are two pomerons. It
is a good working principle always first to try the simplest assumptions, so
we assume that each is a simple pole in the complex angular momentum plane, 
making the contribution from each a simple power of $W$, multiplied by an
unknown function of $Q^2$. Thus
$$
F_2(x,Q^2)=\sum _i f_i(Q^2)\, x^{-\e_i}
\eqno(1)
$$
Bearing in mind that the cross section for the
absorption of real photons is
$$
\sigma ^{\gamma p}={4\pi^2 \alpha _{\hbox{\sevenrm EM}}\over Q^2}F_2\Big\arrowvert
_{Q^2=0}
\eqno(2)
$$
we require each $F_2(x,Q^2)$ to vanish linearly with $Q^2$ at fixed $W$, so that
presumably $f_i(Q^2)$ vanishes like $(Q^2)^{1+\e_i}$. Beyond the
fact that, presumably, they are smooth functions of $Q^2$, this is
the only information that we have about the form of the $f_i(Q^2)$. 
In the past\defref\diffdis{
A Donnachie and P V Landshoff, Nuclear Physics B244 (1984)
}\ref{\small}, we used the simplest possible forms:
$$
f_i(Q^2)= A_i \left ({Q^2\over Q^2+a_i}\right )^{1+\e _i} 
\eqno(3)
$$
and just two terms, with powers
$$\eqalignno{
\e_1=0.0808 &~~~~~~~~~~~~\hbox{(soft pomeron exchange)}\cr
\e_2=-0.4525&~~~~~~~~~~~~\hbox{($f,a$ exchange)}
&(4)\cr}
$$
Now, we retain the same two powers $\e_1$ and $\e_2$, and we add a third term
($i=0$), which we call  hard-pomeron exchange.

As we have said,
Regge theory is supposed to be applicable if $W^2$ is much greater than all
the other variables.
Thus, we expect it to be valid when $x$ is small enough,
whatever the value of $Q^2$. Just what is meant by small enough is open to
debate: we shall assume, as we have before\ref{\small}, that it means 
$$
x<0.07
$$
and, in the case of the real-photon data, $\surd s>6$ GeV,
though it is worth remembering that these choices are somewhat arbitrary.

Because Regge theory gives us no information about the $Q^2$
dependence of the various contributions, we should not expect our simple guess
(3) for the coefficient functions $f_i(Q^2)$ to work for arbitrarily large
$Q^2$. So initially, again as before\ref{\small}, we used it to extract
a preliminary value for the power $\e_0$ by  applying it to 
all data for which
$$
Q^2< 10 \hbox { GeV}^2
$$
including $Q^2=0$. We then abandoned the simple form (3), and used
this preliminary value for $\e_0$ to fit at each value of $Q^2$, and
so found how the data require the coefficient functions $f_i(Q^2)$
to behave at all available values of $Q^2$. This suggested how we should
parametrise the functions $f_i(Q^2)$, and with these parametrisations
we then began the fitting exercise afresh.

The simple form (3) gave $\chi ^2=1.075$ per data point\footnote{$^*$}%
{In our calculations of $\chi ^2$ we have folded
the systematic and statistical errors in quadrature.}, for the 280 data points
with $Q^2<10$, with parameter values 
$$\eqalignno{
~~~~~~~~~~~~~~~~~~~~~~~~~~~~~~~~~~~~~~~~\e_0=0.435~~~~~~~~~~~~~~~~~~~~~\cr
A_0=0.0358~~~~~~~~~~~~~~~~~~~~~~~~~~~~~~~~~~~~~~~~~~~~~~~~&a_0=5.41~~~~~~~~~~~~~~~~~~~~~~~~~~~~~~~~\cr
A_1=0.274~~~~~~~~~~~~~~~~~~~~~~~~~~~~~~~~~~~~~~~~~~~~~~~~~&a_1=0.504~~~~~~~~~~~~~~~~~~~~~~~~~~~~~~~~\cr
A_2=0.149~~~~~~~~~~~~~~~~~~~~~~~~~~~~~~~~~~~~~~~~~~~~~~~~~&a_2=0.0196~~~~~~~~~~~~~~~~~~~~~~~~~~~~~~~&(5)\cr}
$$
We have given each parameter to high accuracy because
a small change in any one parameter has a significant effect on the $\chi ^2$
of the fit. This does not mean that the parameters are  determined to that
accuracy, because the $\chi ^2$ minimum is not very sharp and
a change in any one parameter can be compensated by changes in all the
others in such a way as not to have much effect on the $\chi ^2$.

\input epsf
\topinsert
\line{{\epsfxsize=0.48\hsize\epsfbox{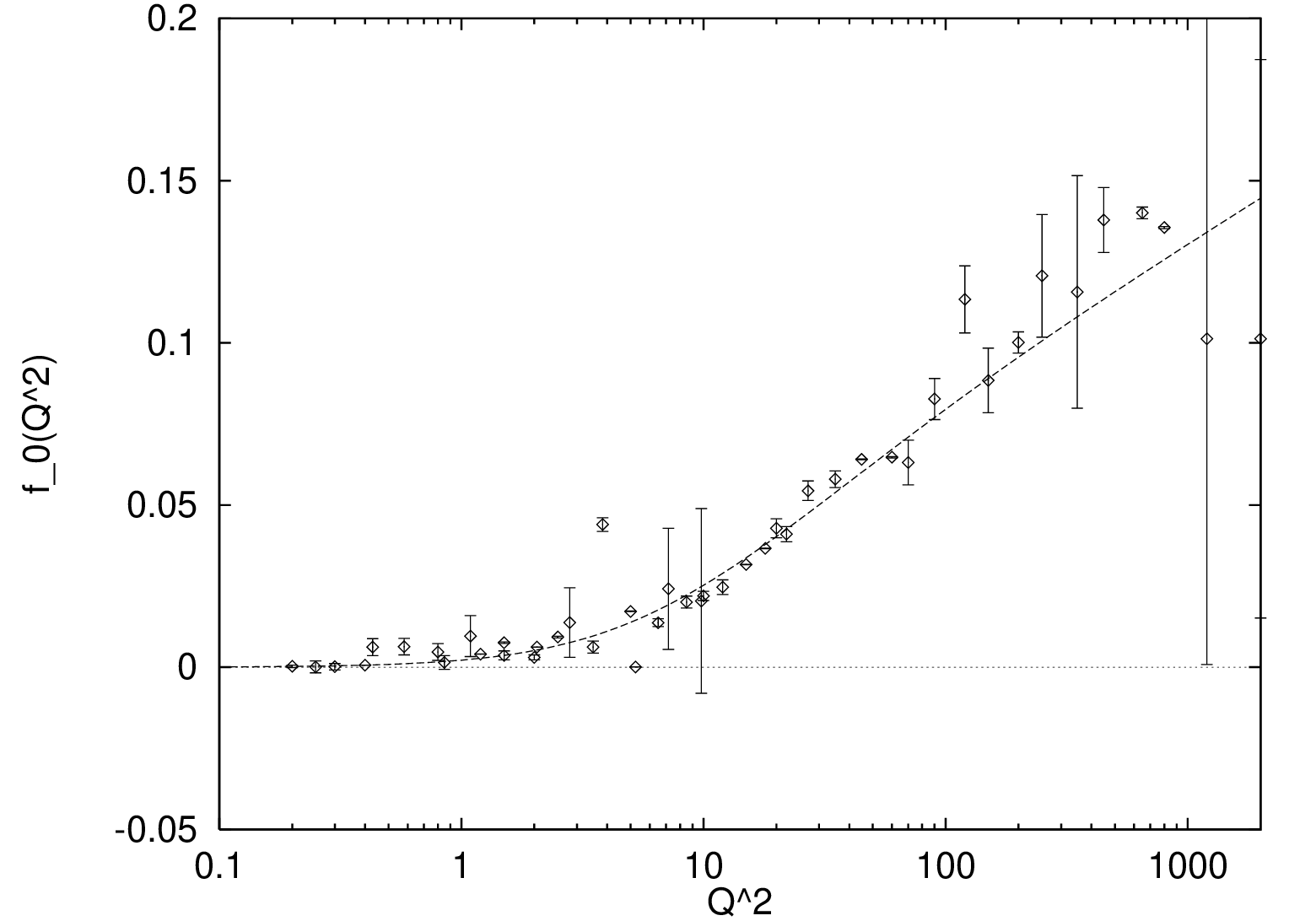}}
{\epsfxsize=0.48\hsize\epsfbox{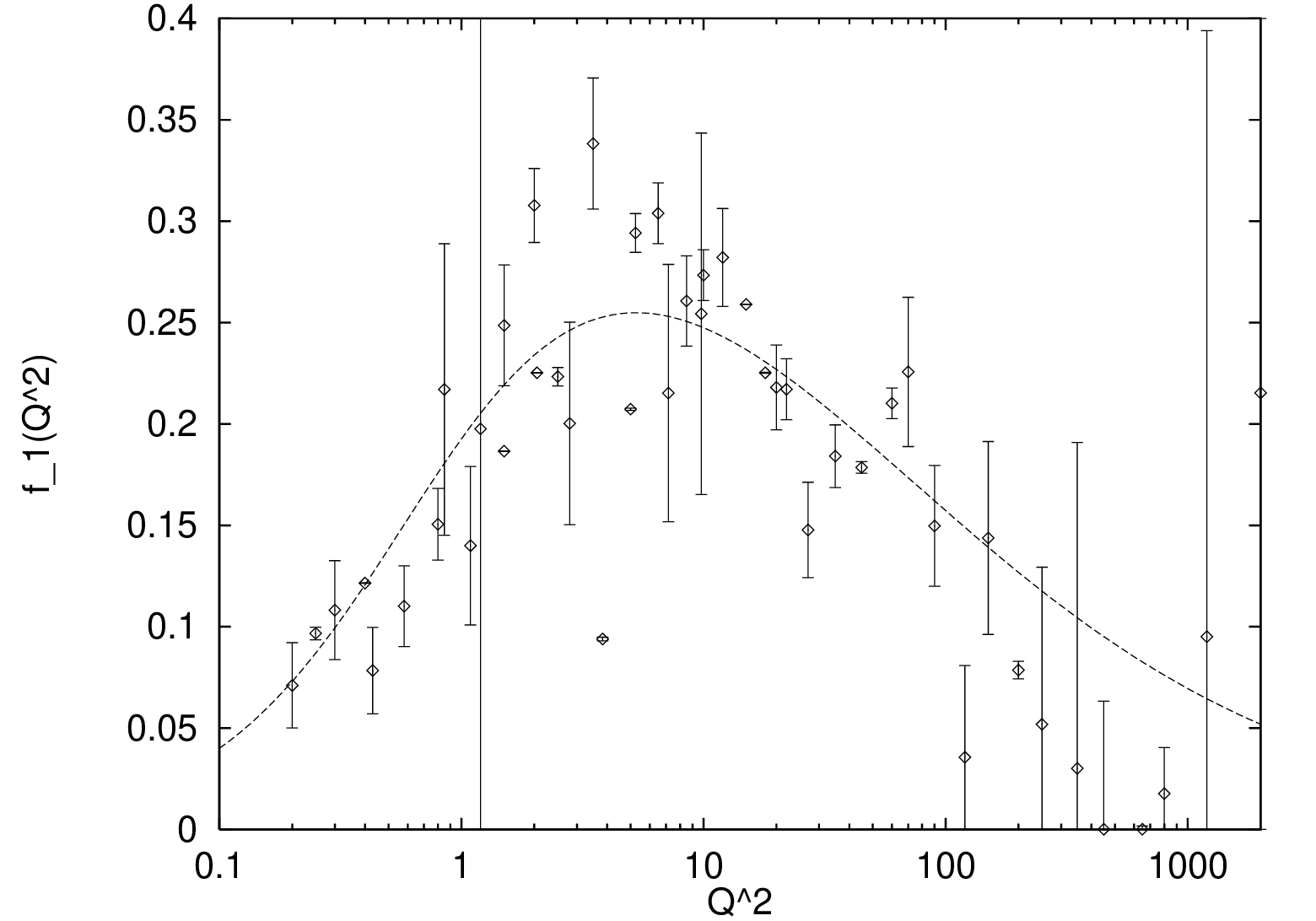}}}\h
Figure 1: Best fits to the coefficient functions $f_0(Q^2)$ and $f_1(Q^2)$,
at each $Q^2$ for which there are data with $x<0.07$, together with curves
corresponding to the fit (6) and (7).
\endinsert

For larger $Q^2$, we find that the simple $Q^2$ dependences make the 
coefficient function of the $x^{-\e_0}$ term
too small, and that of $x^{-\e_1}$ too large. For each value of $Q^2$
for which we have data with $x<0.07$, we made a best fit using the form (1)
keeping the same simple shape (3) for the $f, a$ contribution
and with the values of all the quantities except $f_0(Q^2)$ and $f_1(Q^2)$
taken over from the fit to the $Q^2<10$ data. The resulting values for
$f_0(Q^2)$ and $f_1(Q^2)$ are shown in figure 1, where the error bars
are produced from MINUIT. 
Obviously one can fit these with a variety of functions. We have tried several.
Our lowest overall $\chi ^2$ corresponds to the choices
$$
f_0(Q^2)=A_0 \left ({Q^2\over Q^2+a_0}\right )^{1+\e _0}
\left ( 1+X\log\left (1+{Q^2 \over Q_0^2}\right )\right  ) $$$$
f_1(Q^2)=A_1 \left ({Q^2\over Q^2+a_1}\right )^{1+\e _1}
 {1\over 1+\sqrt{Q^2/Q_1^2}}
$$$$
f_2(Q^2)=A_2 \left ({Q^2\over Q^2+a_2}\right )^{1+\e _2}
\eqno(6)
$$
We vary all the parameters except $\e_1$ and $\e_2$. This 10-parameter fit
gave a minimum $\chi ^2=1.016$  per data point for 539 data points
with the parameter values\goodbreak
$$\eqalignno{
~~~~~~~~~~~~~~~~~~~~~~~~~~~~~~~~~~~~~~~~\e_0=0.418~~~~~~~~~~~~~~~~~~~~~\cr
A_0=0.0410~~~~~~~~~~~~~~~~~~~~~~~~~~~~~~~~a_0&=7.13~~~~~~~~~~~~~~~~~~~~~~~~~~~~~~~~~~~~~~~~\cr
X&=0.485~~~~~~~~~~~~~~~~ Q_0^2=10.6\cr
A_1=0.387~~~~~~~~~~~~~~~~~~~~~~~~~~~~~~~~~a_1&=0.684~~~~~~~~~~~~~~~~ Q_1^2=48.0\cr
A_2=0.0504~~~~~~~~~~~~~~~~~~~~~~~~~~~~~~~a_2&=0.00291~~~~~~~~~~~~~~~~~~~~~~~~~~~~~~~~~~~~~~~~&(7)\cr
}
$$
The same comments as before must be made about the accuracies of
these parameters.
Figures 2, 3 and 4 show the fits they give to the data.

\input epsf
\pageinsert
\epsfxsize=\hsize\epsfbox{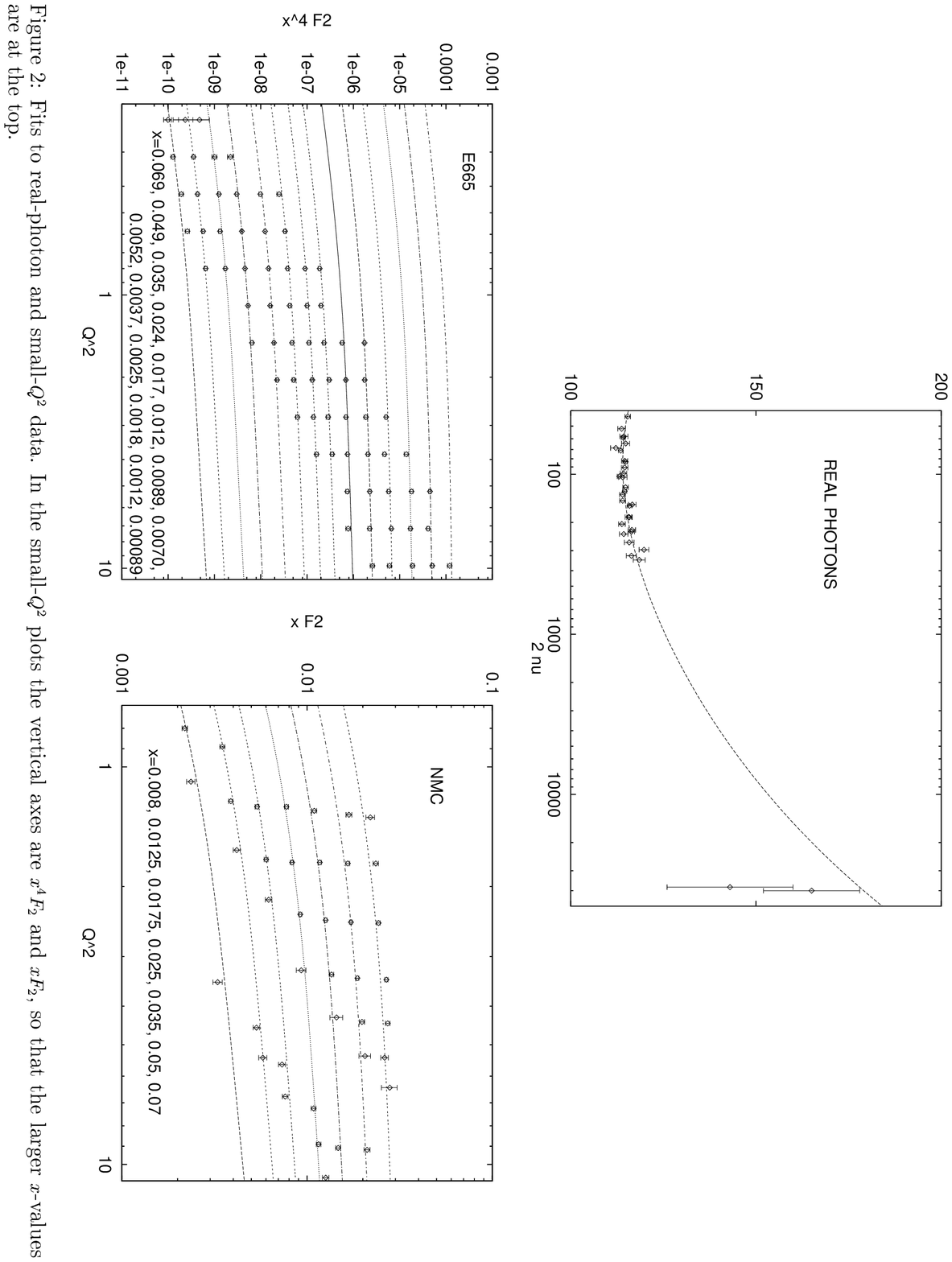}
\endinsert\pageinsert
\epsfxsize=\hsize\epsfbox{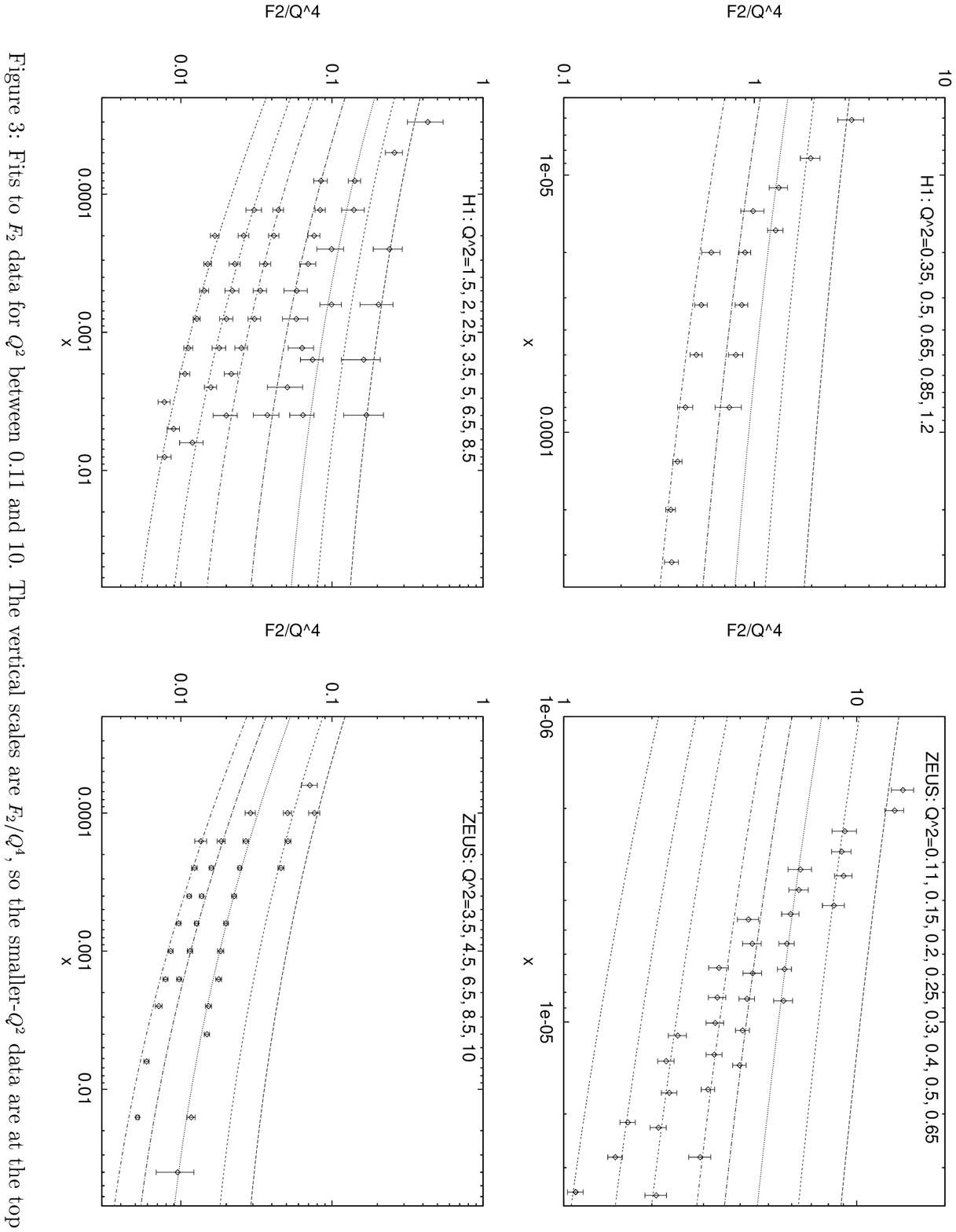}
\endinsert
\pageinsert
\epsfxsize=\hsize\epsfbox{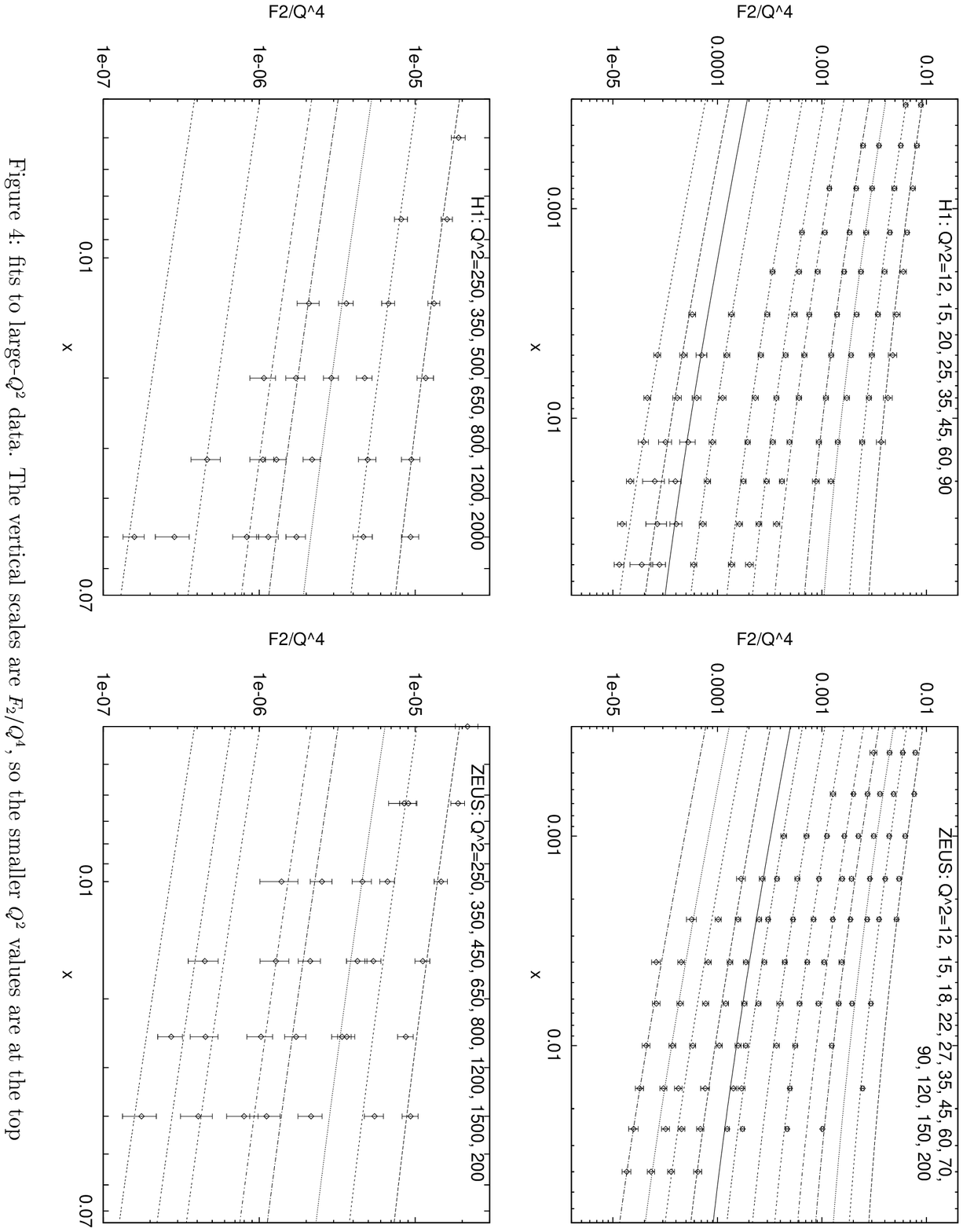}
\endinsert

\bigskip
We have a number of comments to make:

1. Our fits show that,  at small enough $x$, the data may be fitted by
a combination of three fixed powers of $x$. This is in accord with
the Regge-theory expectation if, at least to a good approximation, the
singularities in the complex angular momentum plane are simple poles.
We have used as input the same powers, $\e_1=0.0808$ and $\e_2=-0.4525$, 
for soft pomeron 
and $f,a$ exchange as in our previous fits\ref{\sigtot}, though these
values have recently been questioned by Cudell et al\defref\cudell{
J-R Cudell, Kyungsik Kang and  Sung Ku Kim, talk at 7th Blois Workshop on
Elastic and Diffractive Scattering, hep-ph/9712235
}. Instead of (4), they advocate
$\e_1=0.094$ and $\e_2=-0.33$.
With these values, and the form (6) for the coefficient functions $f_i(Q^2)$,
our $\chi^2$ per data point would reduce to 0.98, with $\e_0=0.410$, though
whether this reduction is significant is doubtful, since our $\chi^2$
is already so close to 1. 

2. There may be a significant
error on our output hard-pomeron power $\e_0=$0.418,
perhaps $\pm 0.05$. For example, if one were to decide that the value 
$a_2=(53.9$ MeV$)^2$ in (7) is too small to be physically reasonable, then
increasing it to the square of the pion mass increases the
$\chi ^2$ per data point to 1.032, and so still gives a good fit, but it reduces
$\e_0$ to 0.388. Alternatively, if we prefer to have the coefficient function
$f_1(Q^2)$ for the soft pomeron go to zero at large $Q^2$ like 1/$Q^2$ instead
of $1/Q$, so that we replace $\sqrt{Q^2/Q_1^2}$ in (6) with 
${Q^2/Q_1^2}$, the $\chi ^2$ per data point becomes 1.018 with $\e_0=0.462$.

\topinsert
\line{{\epsfxsize=0.5\hsize\epsfbox{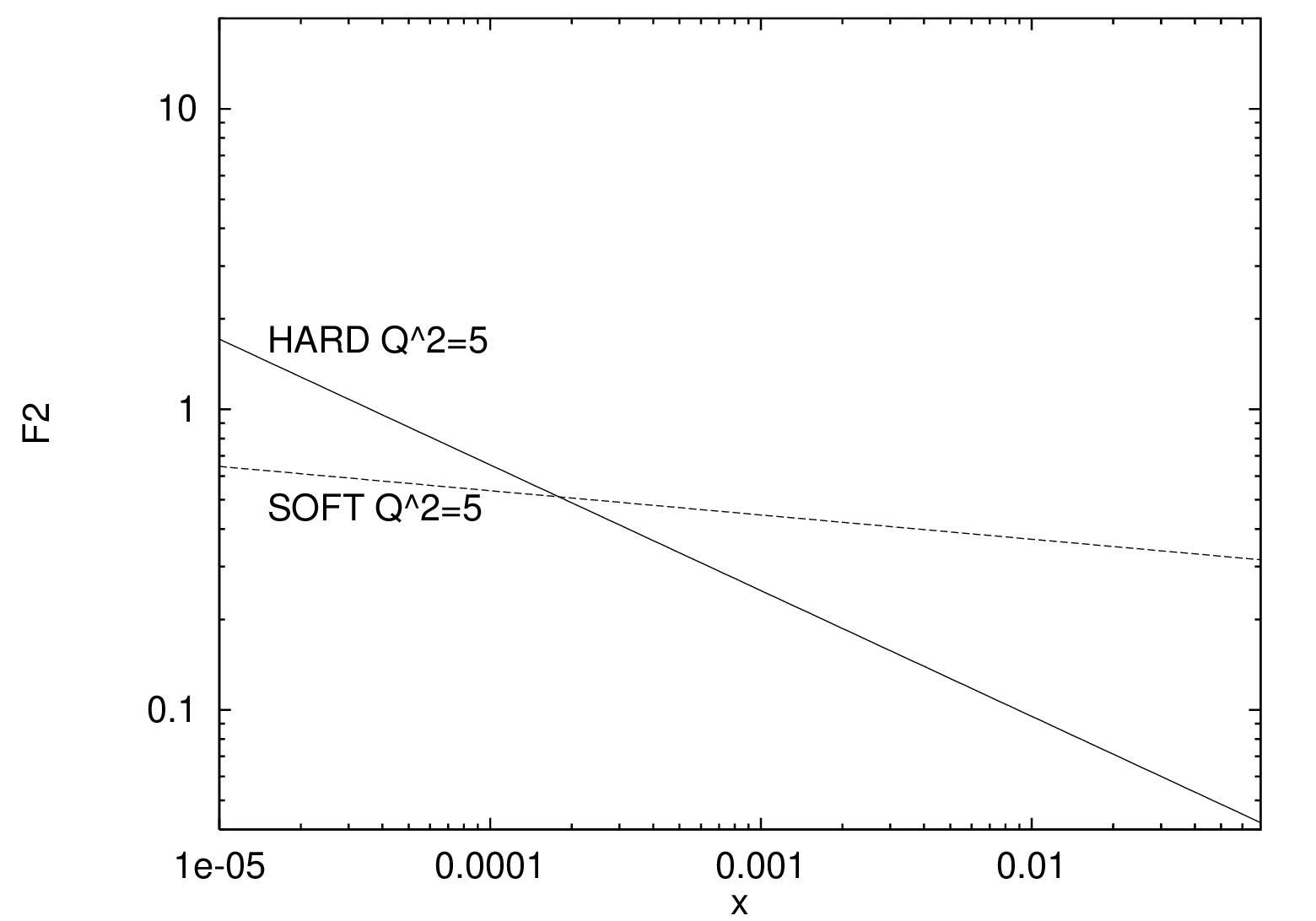}}
\hfill{\epsfxsize=0.5\hsize\epsfbox{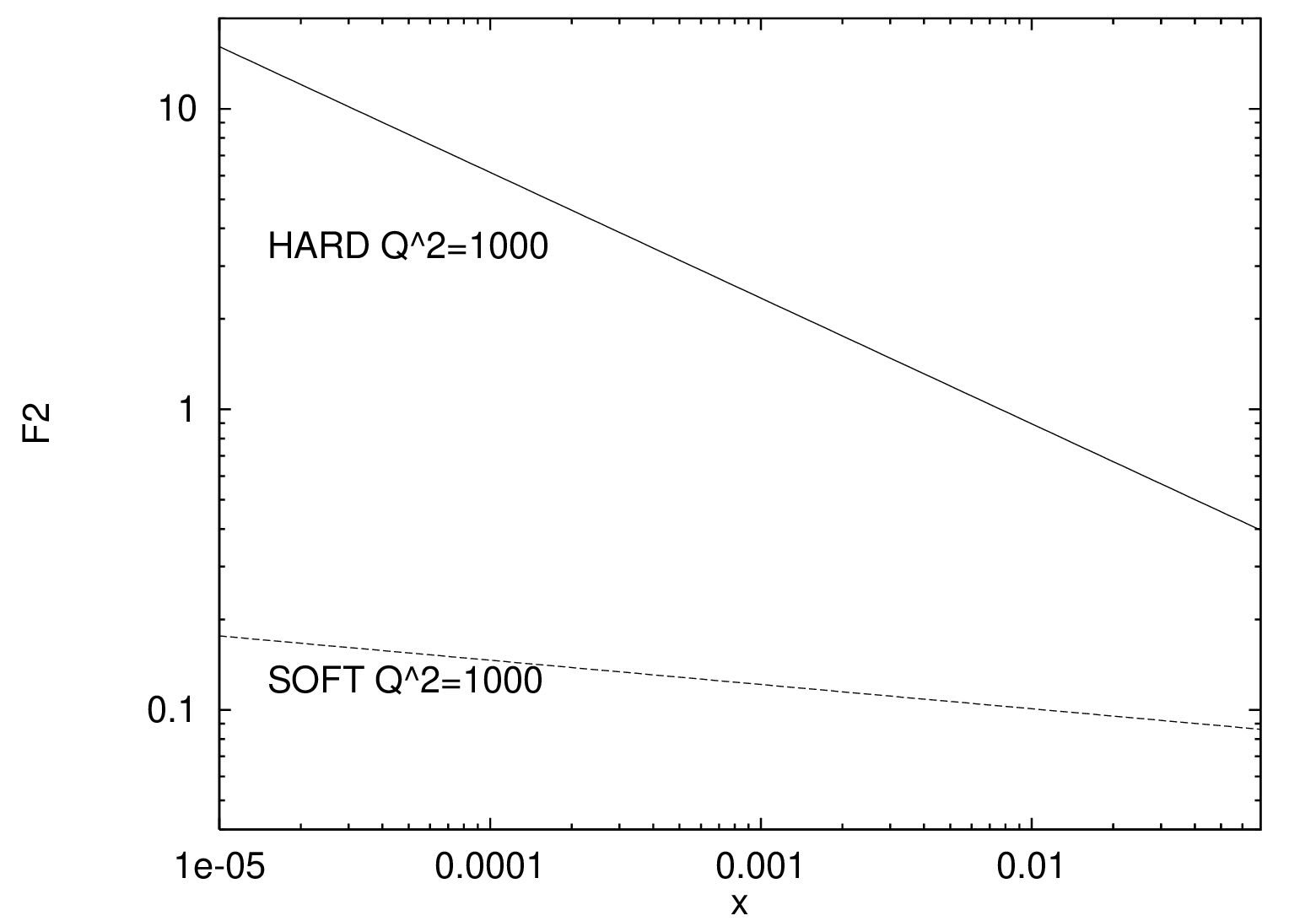}}}

\centerline{Figure 5: Relative magnitudes of hard and soft pomeron 
contributions at $Q^2=5$ and $1000$}
\endinsert

3. This makes it clear that the exact
forms we have used for the $Q^2$ dependences of the three contributions
should not be taken too seriously: Regge theory has nothing to say about
them and we have chosen the simplest forms that work. Nevertheless, we
can be reasonably confident that, as is seen in figure 1, 
the soft-pomeron term has a coefficient that
rises to a maximum somewhere between $Q^2=5$ and 10, and it is very likely that it
goes to zero at high $Q^2$. The hard and soft pomeron contributions to
$F_2$ are shown in figure 5 at two values of $Q^2$; they move in opposite
directions as $Q^2$ increases, so that while the soft pomeron
is important at $Q^2=5$ it becomes less so at higher $Q^2$.
The conclusion that the soft pomeron is higher
twist is one that we did not expect and implies that
all fits based on conventional evolution that have been made
so far need to be reassessed.
It is reminiscent
of a suggestion of McDermott and Forshaw\defref\mf{
M F McDermott and J R  Forshaw, Nuclear Physics B484 (1997) 283
},
who indeed have a next-to-leading contribution that goes to zero as $1/Q$
at high $Q^2$. Their leading term rises like $Q$ at high $Q^2$; the data
are not inconsistent with this: we find it gives a $\chi ^2$ per data point
of 1.072, with $\e_0=0.385$. But our logarithmic form, which presumably 
corresponds more to DGLAP evolution, gives a better fit -- though we emphasise
that Regge theory requires that the perturbative evolution should be
applied in such a way that, at small $x$, it does not cause the power
$\e_0$ to change with $Q^2$.

4. We have said that our choice $x<0.07$ for the range of $x$ in which we
fit the data is arbitrary. Indeed, one might be surprised that the Regge
region extends to such a large value of $x$. If we use instead only data
for which $x<0.01$, with the functional forms (6) for
the coefficient functions $f_i(Q^2)$ and the parameter values (7),
the $\chi^2$ improves from 1.016 per data point to 0.95 (with 357 data points).

5. There will be some who will object to fixed powers of $x$ because of
worries about unitarity. They will be wrong to do so, for several reasons. 
Firstly, we have shown that the fixed powers fit the data, and the data
certainly respect whatever constraints unitarity may impose. Secondly,
unitarity has nothing useful to say about structure functions anyway: the
derivation of the unitarity constraints on a purely hadronic process
such as $pp$ scattering relies on the fact
that a $pp$ intermediate state appears in the
unitarity equation; but $\gamma^*p$ scattering is very different, because
there is no $\gamma^*p$ intermediate state. In any case, even in $pp$ 
scattering,  data at present energies are not seriously affected by unitarity
constraints. 
The fact that a constant power fits the data well\ref{\sigtot} indicates
that, so far, shadowing corrections to the $t=0$ amplitude are small, though
they will become more important at higher energies. Similarly, 
at extremely small $x$ the fixed-power behaviour
will be moderated by shadowing suppression, but there is no reason to
believe that such small values have yet been achieved.

6. The accurate data for the low-energy real-photon cross-sections are
an important constraint. If we make a best fit to the virtual-photon data
alone, its extrapolation comes nowhere near the low-energy real-photon
data.  
As can be seen in figure 2, our new 
fit extrapolates the small-$Q^2$ data to give a larger real-photon
cross-section at HERA energies than in our old fits\ref{\sigtot}. The ZEUS
collaboration has reached a similar conclusion\defref\doyle{
ZEUS: talk by A T Doyle at DIS98, Brussels (1998)
}, but it must be emphasised that it depends crucially on assuming a
simple form for the $Q^2$ dependence. There is no reason to rely on this,
and by adopting more complicated extrapolations one can arrive at lower 
values\defref\allm{
H Abramowicz and  A Levy, DESY 97-251\h
G Kerley and G Shaw, Physical Review D56 (1997) 7291
}. We could even decouple the hard pomeron completely at $Q^2=0$.
Nevertheless, it is interesting to note that preliminary results from
a new ZEUS measurement of the real-photon total cross-section agree rather
well with our curve\defref\surrow{
B Surrow, DESY-THESIS-1998-004
}, 
which is
$$\eqalignno{
\sigma ^{\gamma p}&=4\pi^2\alpha_{\hbox{{\sevenrm EM}}}\sum_i A_i\,a_i^{-1-\e_i}
(2\nu)^{\e_i}\cr
&=0.283\,(2\nu)^{0.418}+65.4\,(2\nu)^{0.0808}+138\,(2\nu)^{-0.4525}&
(9)\cr}
$$
with $2\nu$ in GeV$^2$ and the coefficients in $\mu$b, where $2\nu=W^2-m_p^2$.

\pageinsert
\epsfxsize=\hsize\epsfbox{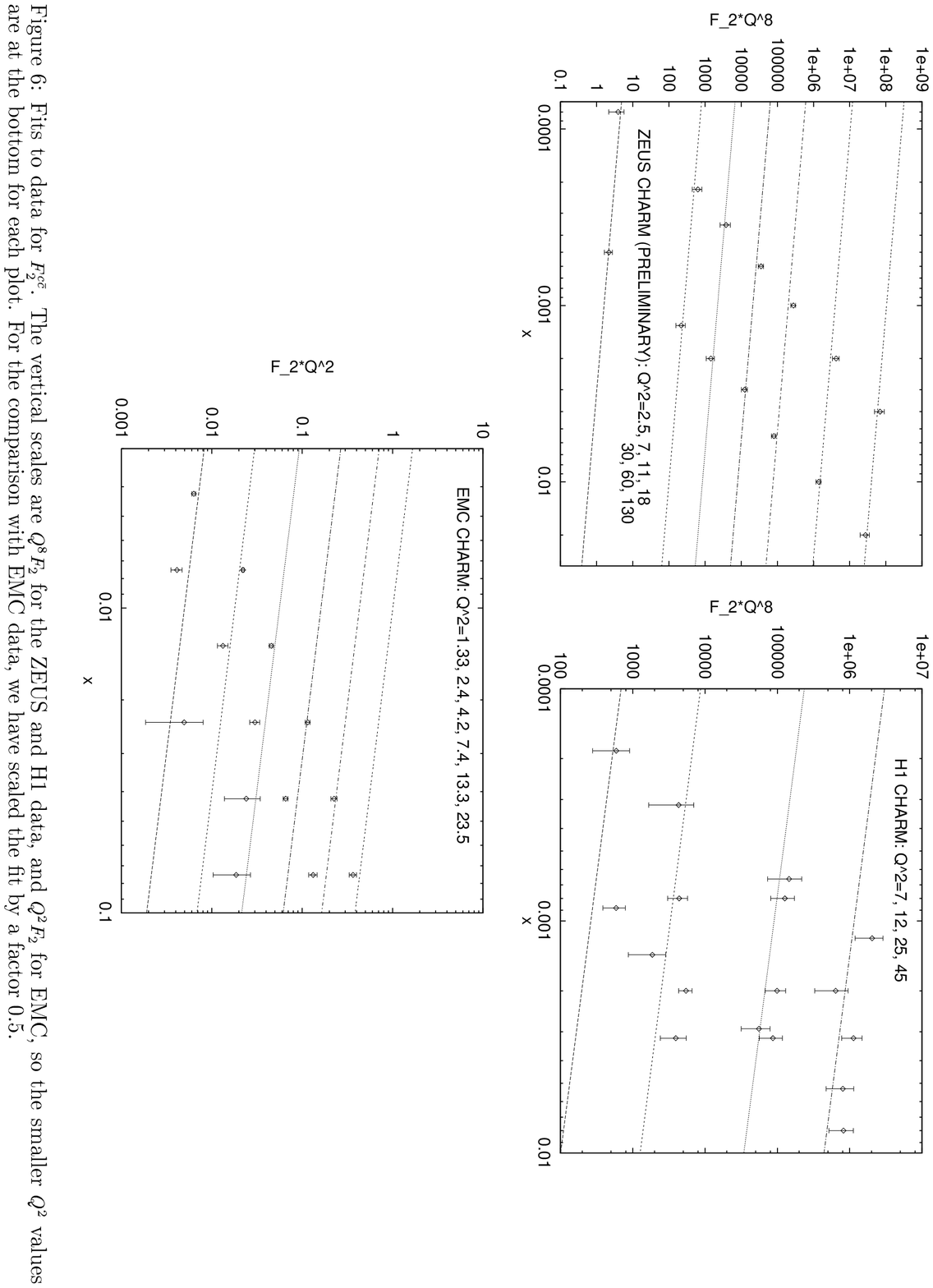}
\endinsert

7. We have made the assumption that the 
$Q^2$ dependences of the contributions to $F_2$ from different flavours
are the same, or at least that when the contributions are added together
their sum has simple behaviour. This is not obviously true, and
it is interesting to subject the recent
accurate preliminary ZEUS data\ref{\doyle}
for the charm content of $F_2$ to our parametrisation.
We find that they indicate that both the soft-pomeron and the
$f,a$ exchanges do not couple. The latter is readily understandable
because of Zweig's rule, but the decoupling of the soft pomeron comes as
something of a surprise. We impose
$$
F_2^{c\bar c}=A_{\hbox{{\sevenrm C}}} \left ({Q^2\over Q^2+a_{\hbox{{\sevenrm C}}}}\right )^{1+\e _0}
\left ( 1+X_{\hbox{{\sevenrm C}}}\log(Q^2/Q_{\hbox{{\sevenrm C}}}^2)\right  )
x^{-\e_0}
\eqno(10)
$$
with $\e_0=0.418$ as before. The data are not sufficient to determine the remaining
four parameters well, so we choose to impose a constraint that the coupling of
the hard pomeron is flavour-blind at large enough $Q^2$. Assuming that
so far $Q^2$ is not large enough for there to be a significant contribution
from $b$ quarks, this implies that the coefficient of the logarithm in
the hard-pomeron coefficient function for 
$F_2^{c\bar c}$ is
\hbox{$\textstyle {4\over 9}/({4\over 9}+{1\over 9}+{1\over 9}+{4\over 9})$}
times that for the complete $F_2$:
$$
A_{\hbox{{\sevenrm C}}}X_{\hbox{{\sevenrm C}}}=0.4\,A_0X
\eqno(11a)
$$
Then, the 3-parameter fit to the ZEUS data yields
$$
A_{\hbox{{\sevenrm C}}}=0.184~~~~~~
a_{\hbox{{\sevenrm C}}}=8.45~~~~~~
X_{\hbox{{\sevenrm C}}}=0.431~~~~~~Q_{\hbox{{\sevenrm C}}}^2=212
\eqno(11b)
$$
and is shown in figure 6. 
This figure also includes comparison of the 
fit to the earlier H1 and EMC data\defref\emc{
EMC: J J Aubert et al, Nuclear Physics B213 (1983) 31\h
H1:  C Adloff et al, Z Physik C72 (1996) 593; J. Breitweg et al,
Physics Letters B407 (1997) 402
}. 
In the case of the EMC data, we have had to renormalise our fit with a constant
factor 0.5 to get good agreement. We cannot understand why this should be,
but the latest
MRS fit\defref\mrs{
A D Martin, R G Roberts, W J Stirling and R S Thorne, hep-ph/9803445 
}
similarly is twice as large as the low-$x$ EMC charm data.
Ideally, one would first make a fit to the charm-quark data, then subtract it
from the complete $F_2$ and make another fit to just the light-quark
contribution, but this must wait until the data improve. Eventually, the charm
data will be important for reducing the error on the value of the hard-pomeron
power $\e_0$: taken literally, the present charm data suggest that it should be
towards the upper end of the range $0.42\pm 0.05$ that we have quoted above.

8. We have found that we need a ``hard-pomeron'' term, with Regge intercept
about 1.4. It is perhaps ironic that we reach this conclusion just as
the hard BFKL pomeron seems to be in retreat\defref\ross{
V S Fadin and L N Lipatov, hep-ph/9802290\h
M Ciafaloni and G Camici, hep-ph/9803389\h
D A Ross, hep-ph/9804332
}. This does not rule out the possibility that our hard pomeron is
perturbative. After all --- as we have consistently maintained\defref\bkf{
J C Collins and P V Landshoff, Physics Letters B276 (1992) 196\h
J R Cudell and A Donnachie and P V Landshoff, Nuclear Physics B482 (1996) 241
}$^,$\ref{\mf}
--- in the BFKL context asymptopia is very far away from present physics.
Whatever the explanation of the hard pomeron, it is interesting that the mass
scale $a_0$ that determines how rapidly its contribution to $F_2$ rises
with $Q^2$ is considerably larger than the corresponding soft-pomeron scale 
$a_1$.
Nevertheless, it may well be that even the hard pomeron is nonperturbative
and that it is just a glueball trajectory. There is already some experimental
support for the notion that this is true for the soft pomeron: the $2^{++}$
glueball candidate\defref\wa{
WA91: S Abatzis et al, Physics Letters B324 (1994) 509
}
at 1926 MeV
is at just the right mass to lie on a straight trajectory of intercept close
to 1.08 and slope 0.25 GeV$^{-2}$. Another $2^{++}$ glueball candidate
has just been announced\defref\barberis{
WA102:  D. Barberis et al, hep-ex/9805018
},
with a mass 2350 MeV. Let us suppose that this lies on the hard pomeron
trajectory, and that this is also straight. The hard and soft pomeron
trajectories will apparently
then intersect in the complex angular momentum plane; 
however it is likely\footnote{$^*$}{We are grateful to Peter Collins for a
correspondence about this.} 
that mixing will rather cause them to avoid each other in the way shown in 
figure 7. This picture gives the hard-pomeron trajectory a slope
close to 0.1  --- though we caution that figure 7 should not be taken too
seriously.
\topinsert
\centerline{{\epsfxsize=0.4\hsize\epsfbox{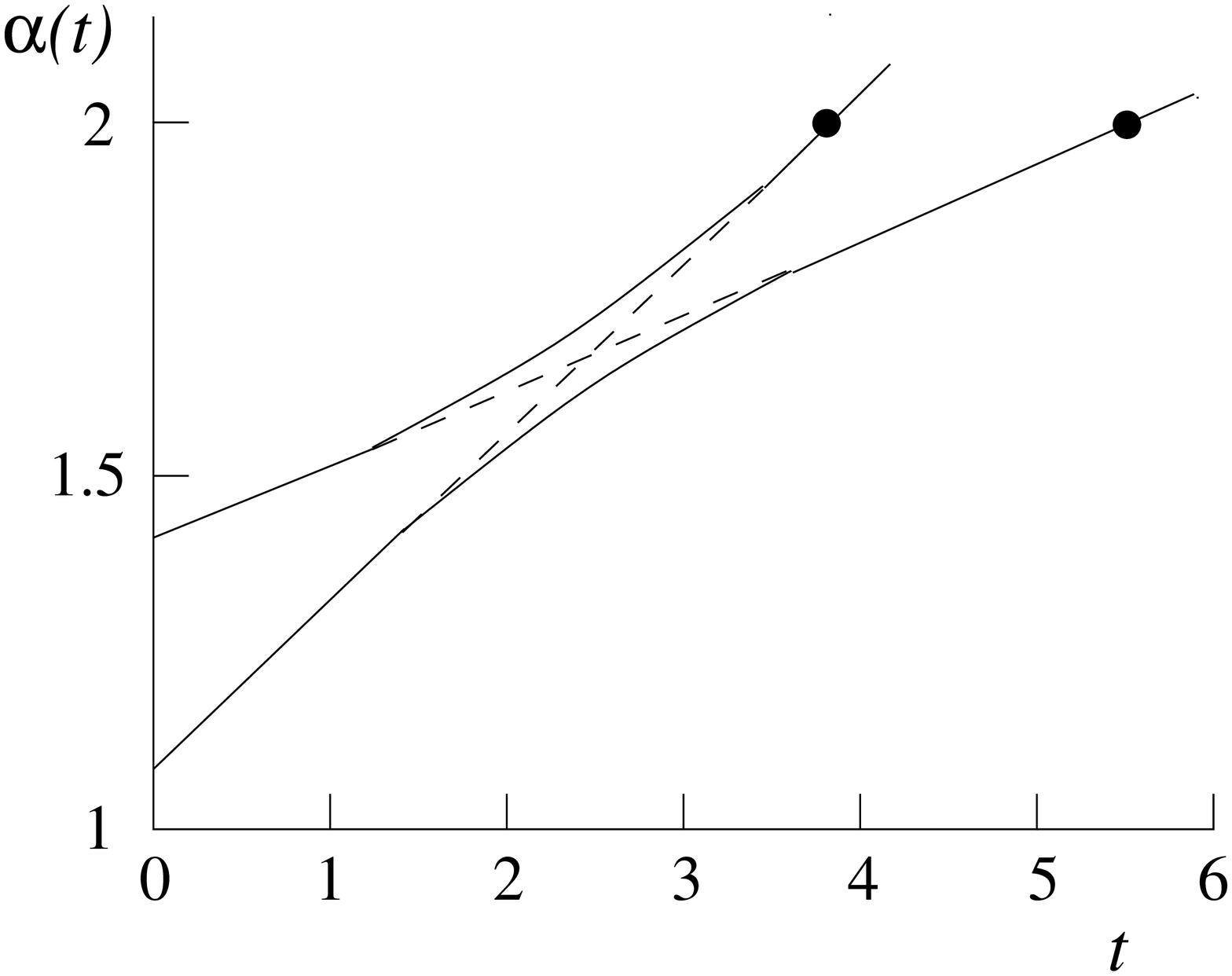}}}

\centerline{Figure 7: Conjectured
soft and hard pomeron trajectories, with 2$^{++}$ 
glueball candidates\ref{\wa}\ref{\barberis}}
\endinsert

9. Although data for quasi-elastic $\rho$ photoproduction
are well described by soft pomeron exchange, this is not the case for
$J/\psi$ photoproduction. Not only is there a much more 
rapid rise with increasing
energy\defref\jpsi{
ZEUS: J Breitweg et al, Z Physik C75 (1997) 215\h
H1: S Aid et al, Nuclear Physics B472 (1996) 3
}, but also there is much less shrinkage of the forward peak\defref\levy{
A Levy, hep-ph/9712519
}. This second feature suggests that the slope of the hard pomeron's 
trajectory may be small, at least near the forward direction, as we
have indeed shown in figure 7.
Assuming this, the contribution to the amplitude from the exchange of the two
pomerons would be of the form
$$
T(s,t)=i\sum _{i=0}^1 B_i(t)s^{e_i(t)}\,e^{-\half i\pi e_i(t)}
$$$$
e_0(t)=\e _0+0.1 t ~~~~~~~~~~~~~~~~~~~~~~~~~e_1(t)=\e _1+0.25 t
\eqno(12a)
$$
The functions $B_i(t)$ include
products of the form factors that couple the pomerons
to the proton and the $\gamma J/\psi$ vertex. We model them 
with the functions
$$
B_i(t)=B_i(0)e^{2t}
\eqno(12b)
$$
though their exact shape is not critical for our analysis -- it has just
a small effect on the energy dependence.
The best fit to the existing data has 
$$
B_0(0)/B_1(0)=0.05
\eqno(12c)
$$
and is shown in figure 8. It is interesting that H1 has new preliminary data
points at higher energies\defref\hhpsi{
H1: talk by A Bruni at DIS98, Brussels (1998)
} 
which agree well with our fit and confirm that the curve really should be 
concave upwards.

\topinsert
\centerline{{\epsfxsize=95truemm\epsfbox{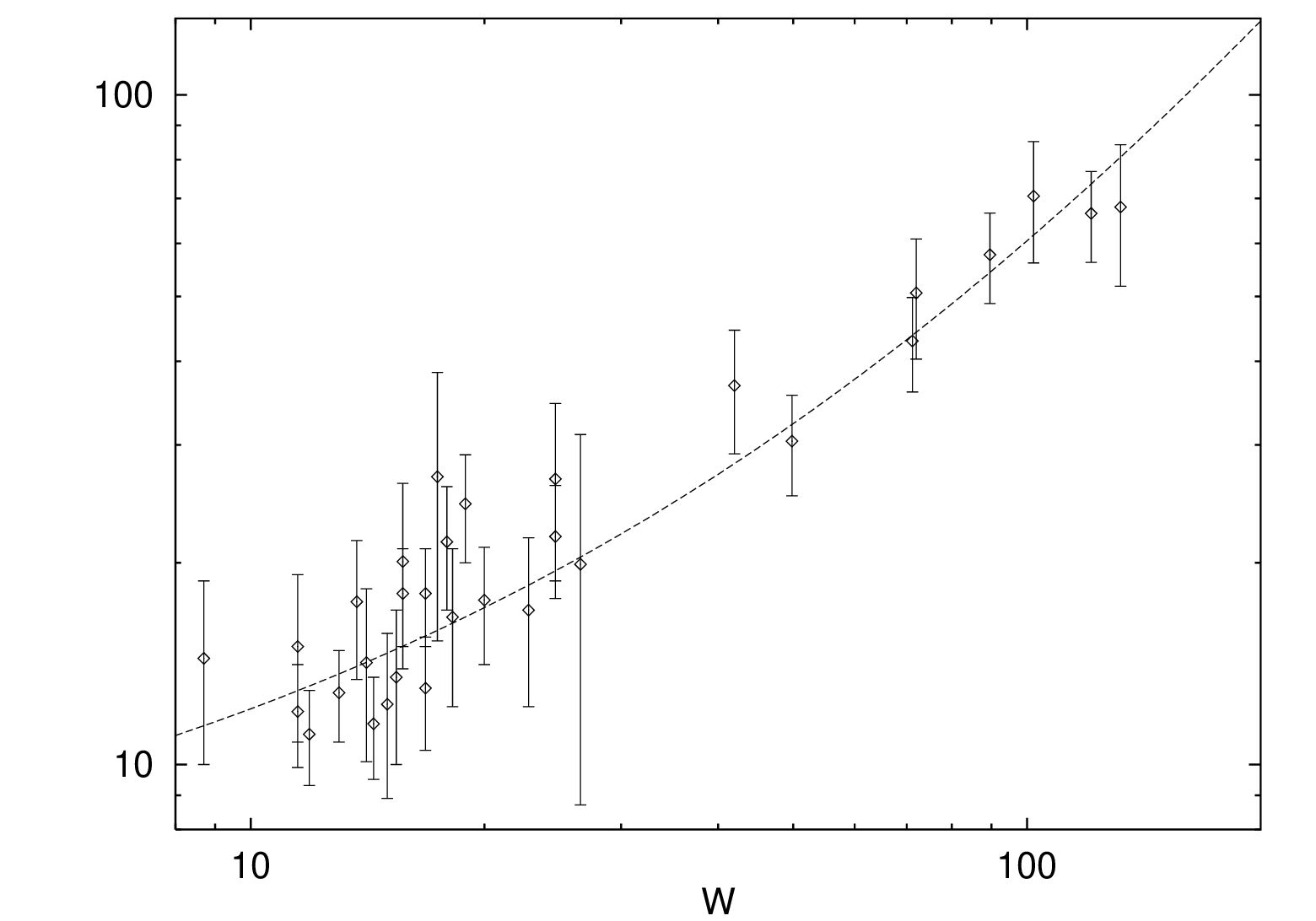}}}
\centerline{$\surd s$}\medskip
\centerline{Figure 8: $\sigma ( \gamma p\to J/\psi\, p)$ in microbarns,
together with simple fit}
\endinsert

10. An obvious question is whether the hard pomeron contributes also
to purely soft processes. There is perhaps a hint of a small contribution
to exclusive $\rho$ photoproduction, in that the $t$-slope seems to be
slightly smaller than what purely soft-pomeron exchange would give\ref{\hhpsi}.
The contribution of the soft pomeron to the $pp$
and $\bar pp$ total cross-sections was already fixed rather 
accurately\defref\oldfit{
A Donnachie and P V Landshoff, Physics Letters 123B (1983) 345
}
by the data up to ISR energies, but there is room for some extra contribution
in the higher-energy data. With some small downwards adjustment of
the soft-pomeron term, one can tolerate a hard-pomeron contribution
of perhaps
$$
0.01\, s^{\e_0}
\eqno(13)
$$
This would give slightly better
agreement with the UA4 measurement than we had before\defref\ua{
M Bozzo et al. Physics Letters 147B (1984) 392
}
but would still
favour the E710 measurement over that of CDF\defref\tev{
E710: N Amos et al, Phys Lett B243 (1990) 158\h
CDF: F Abe et al, Physical Review D50 (1994) 5550
}. 

11. The coefficient of the hard-pomeron term (13) in the $pp$ total cross
section is more than 2000 times smaller than for the soft-pomeron term.
However, if it turns out that (9) is the correct $Q^2\to 0$ extrapolation,
then in the $\gamma p$ total cross section the ratio is an order
of magnitude larger.
This would support the viewpoint of those\defref\ss{
G A Schuler and T Sj\"ostrand, Physics Letters B300 (1993) 169
}
who have maintained that the $\gamma p$ total cross section contains a
significant piece that is different in character from purely hadronic cross
sections. This additional piece cannot factorise in the Regge sense, because
it would give a $\gamma\gamma$ total cross section much larger than
has been measured\defref\gg{
S S\"oldner-Rembold, hep-ex/9805008
}. However, the preliminary OPAL and L3 data do hold the prospect that the hard
pomeron may be visible in $\gamma\gamma$ collisions.

12. The main message of this paper has been that Regge theory is applicable
even at very large $Q^2$, if only $x$ is small enough. In this, our work
does not relate immediately to any of the many other fits in the literature:
we insist that the hard pomeron is present already at small $Q^2$ and that
$Q^2$ evolution keeps both it and the other Regge powers intact. 
Kerley and Shaw\ref{\allm} also have two pomerons, though the details of
their work are very different. 
Our approach is in the same spirit as that of Haidt\defref\haidt{
D Haidt, Proceedings of the Workshop on DIS at Chicago, April 1997, AIP 1997
},
though the details are again very different.
The work of Nikolaev, Zakharov and Zoller\defref\nzz{
N N Nikolaev, B G Zakharov and V R Zoller, 
JETP Lett. 66 (1997) 138
}
is closest to ours, though they do not extend their fit below $Q^2=1.5$ and
therefore have rather different powers of $x$ from ours. Like us, they find
that the coefficient functions of the nonleading $x$ powers peak somewhere
around $Q^2=10$, though their behaviour at large $Q^2$ is very different
from ours. Indeed, we suggest that the soft pomeron contribution, although large
for $Q^2$ as large as 10 or more, is probably higher twist.

\goodbreak
\bigskip{\eightit
This research is supported in part by the EU Programme
``Training and Mobility of Researchers", Networks
``Hadronic Physics with High Energy Electromagnetic Probes"
(contract FMRX-CT96-0008) and
``Quantum Chromodynamics and the Deep Structure of
Elementary Particles'' (contract FMRX-CT98-0194),
and by PPARC}
\vfill\eject
\goodbreak
\medskip\immediate\closeout\rfile\writestoppt
\baselineskip=12pt\parskip=4pt{{\bf References}}\bigskip{\frenchspacing%
\parindent=20pt\escapechar=` \input refs.tmp\bigskip}\nonfrenchspacing
\bye